\documentclass[article,twocolumn,amsmath,floats,showpacs,nofootinbib]{revtex4}
\usepackage[usenames]{color}
\usepackage{graphicx}
\usepackage{epstopdf}
\usepackage{textcomp}
\usepackage{hyperref}
\usepackage{multirow}
\usepackage{tikz}
\usepackage{rotating}
\hypersetup{colorlinks=true}

\begin{document}

\title{Point-contact measurements of the energy gap of lead-chalcogenide-based superconducting superlattices}

\author{I.K. Yanson, N.L. Bobrov, L.F. Rybal'chenko, V.V. Fisun, O.A.~Mironov, S.V. Chistyakov, A.Yu. Sipatov, and A.I. Fedorenko}
\affiliation{B.I.~Verkin Institute for Low Temperature Physics and
Engineering, of the National Academy of Sciences
of Ukraine, prospekt Lenina, 47, Kharkov 61103, Ukraine, Institute of Radiophysics and Electronics, Academy of Sciences of the Ukrainian SSR, Kharkov and V.I. Lenin Polytechnical Institute, Kharkov
Email address: bobrov@ilt.kharkov.ua}
\published {(\href{http://www.jetpletters.ac.ru/ps/225/article_3757.pdf}{Pis'ma Zh.Eksp.Teor.Fiz.}, \textbf{49}, No.5, 293 (1989)); (\href{http://www.jetpletters.ac.ru/ps/1116/article_16905.pdf}{JETP Letters}, \textbf{49}, No.5, 335 (1989)}
\date{\today}

\begin{abstract}The energy gap of superconducting $PbTe-PbS$ semiconductor superlattices has been analyzed with the help of point contacts for the first time below the critical temperature, $T_{c}=3.9~K$, and in the fluctuation region $T>T_{c}$. The size of the gap and its dependence on the temperature and field are determined by the position of the point contact inside the superlattice.

\end{abstract}

\maketitle

\begin{figure}[]
\includegraphics[width=8cm,angle=0]{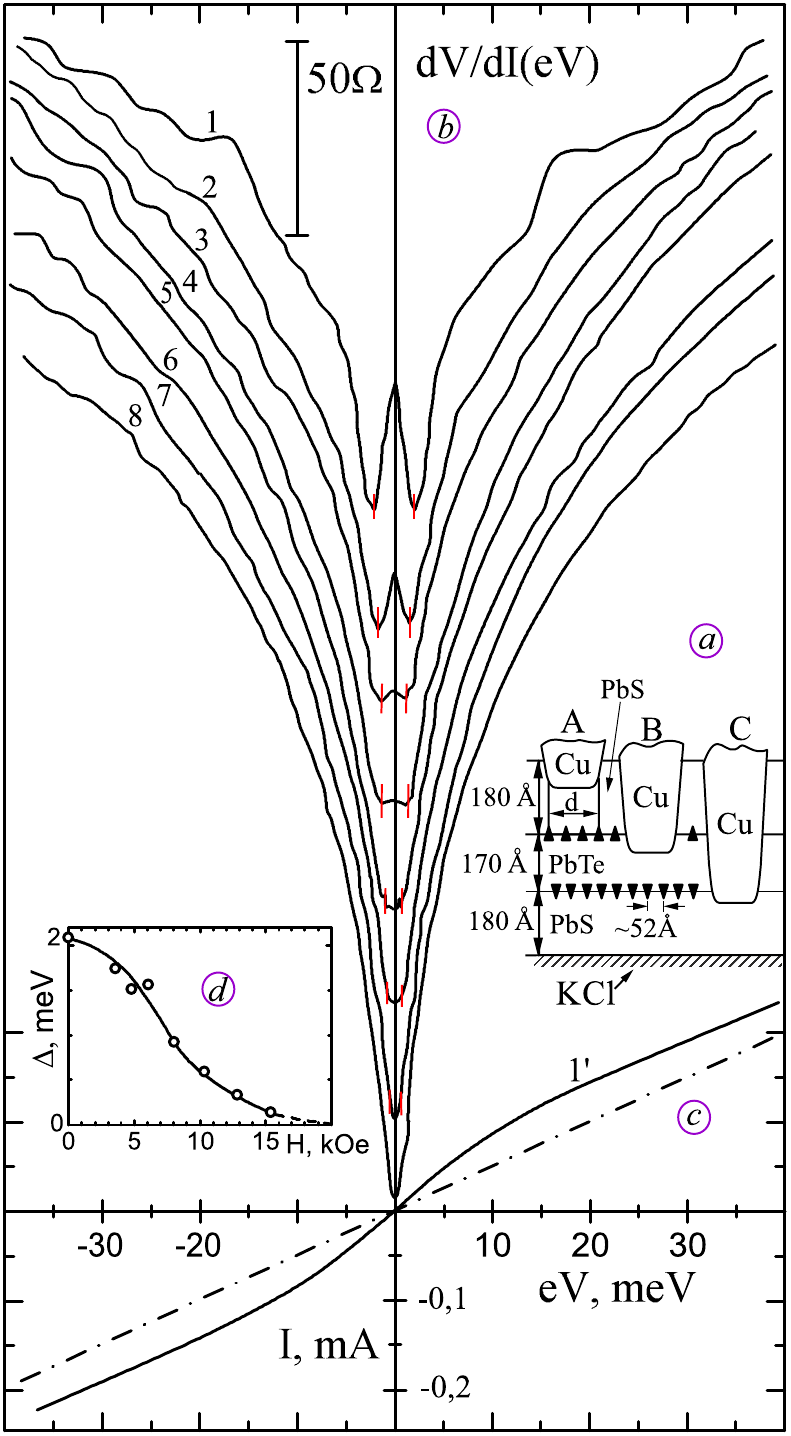}
\caption[]{(a) A three-layer epitaxial structure of $PbS-PbTe-PbS/(001 )KCl$ and various contact models. (b) Differential-resistance curves and (c) the I-V characteristic of a three-layer structure at $T=1.85~Ê$ and $H=0$ (1,1'), 3.8~$kOe$ (2), 5.1~$kOe$ (3), 6.25~$kOe$ (4), 7.9~$kOe$ (5), 10.5~$kOe$ (6), 13~$kOe$ (7), and 15.6~$kOe$ (8). The dashed line is drawn parallel to the I-V characteristic at $eV\gg\Delta$. (d) Energy gap vs the magnetic field.}
\label{Fig1}
\end{figure}

\begin{figure}[]
\includegraphics[width=8cm,angle=0]{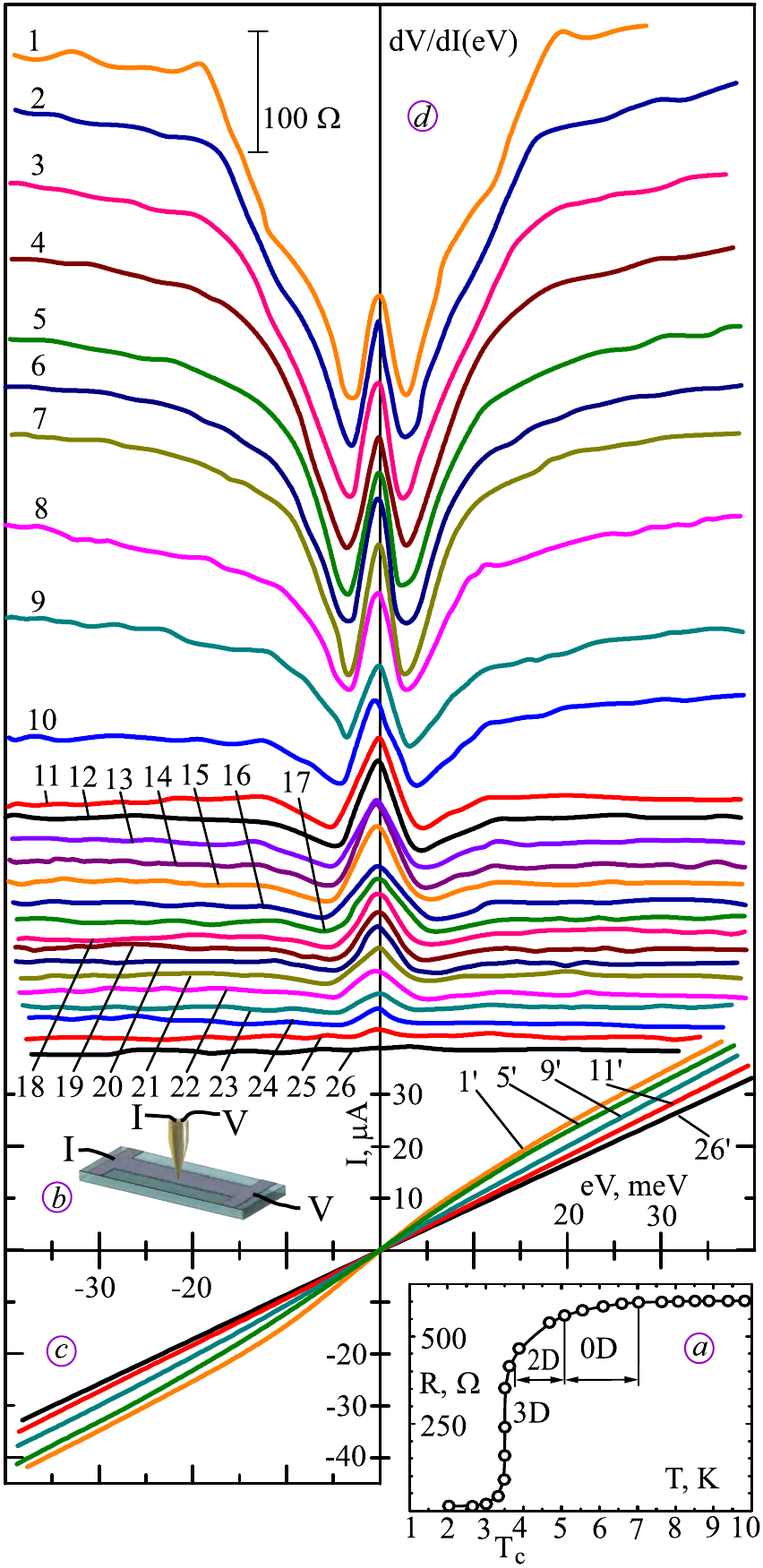}
\caption[]{(a) Temperature dependence of the resistance of the superlattice with $N=1.5$, measured along the direction of the layers. (b) The experimental geometry for the analysis of the I-V characteristic of a point contact of a $PbTe-PbS$ superlattice. Temperature dependence of the I-V characteristics (c) and of the differential resistance (d) for the point contacts of a three-layer epitaxial structure of $PbS-PbTe-PbS/(001)KCl$ at $H=0$ for the following temperatures $T$: 1.85~$K$ (1.1'), 2.01~$K$ (2), 2.2~$K$ (3), 2.36~$K$ (4), 2.6~$K$ (5,5'), 2.8~$K$ (6), 3.0~$K$ (7), 3.28~$K$ (8), 3.6~$K$ (9,9'), 3.8~$K$ (10), 4.0~$K$ (11, 11'), 4.2~$K$ (12), 4.4~$K$ (13), 4.6~$K$ (14), 4.8~$K$(15), 5.0~$K$ (16), 5.2~$K$ (17), 5.4~$K$ (18), 5.6~$K$ (19), 5.8~$K$ (20), 6.0~$K$ (21), 6.2~$K$(22), 6.4~$K$ (23), 6.6~$K$ (24), 6.8~$K$ (25), 7.0~$K$ (26, 26').}
\label{Fig2}
\end{figure}

The epitaxial semiconductor superlattices of $PbTe-PbS$ with structural periods $N=1.5$ and 10 were fabricated by vacuum-technology methods on (001) $KCl$ substrates \cite{1}. In fabricating these superlattices attention is focused principally on the three-layer sandwich $PbS-PbTe-PbS/(001)KCl$ (Fig.\ref{Fig1}a), which is the minimum structural unit ($N=1.5$) that manifests superconducting properties. It has been established that superconductivity in this system has a
 quasi-two-dimensional nature and that it stems from the presence of regular square grids of misfit edge dislocations at the heterojunctions. In the case of a superlattice with $N=10$, a crossover occurs in the plot of $H_{c2}^{\parallel}$ vs the temperature: A transition from a three-dimensional behavior at temperatures $3.2Ê<T<3.8~K$ to a two-dimensional behavior at $T<3.2~K$ occurs as a result of the temperature dependence of the coherent length \cite{1}, $\xi^{\perp}(T)$. At $T=3.2~K$ we have $\xi^{\perp}\approx D$ where $D=35~nm$ is the period of the superlattice, and at $T<2.1~K$ we have $\xi^{\perp}(T)=6-8~nm <D$ and the superconductivity localizes near the grid of the misfit dislocations for any values of $N$. At $T>T_c$ the dimensionality of the superconducting fluctuations can be determined by analyzing the resistive transitions by analogy with Ref. \cite{2}, taking into account the specific features of layered superconductors \cite{3}. At temperatures in the range $4.4~K<T<5~K$ it turned out (see Fig.\ref{Fig2}a) that $\xi^{\perp}=5-10~nm$ and that the temperature dependence of the contribution from the superconducting fluctuations to the conductivity corresponds to a situation which occurs between the two-dimensional case and the three-dimensional case. At higher temperatures, $5~K<T<7~K$ we have $\xi^{\perp}<5~nm$ and the superconducting fluctuations are of a zero-dimensional nature: The superconductivity localizes in the nodes of a grid of misfit dislocations.

Point contacts were created between the point of a copper single-crystal pyramid and the superlattice, directed perpendicular to the layers (Fig.\ref{Fig2}b). To conduct a thorough analysis, we chose point contacts with current-voltage characteristics whose first derivative $dV/dI$ has minima (Figs.\ref{Fig1}b and \ref{Fig2}d) near $V=0$ ($V$ is the voltage across a point contact) distributed in a symmetric arrangement relative to the ordinate. Such minima usually are attributable to the manifestation of an energy gap $\Delta$ on
the I-V characteristics of $ScN$ and $ScS$ junctions with dimensions smaller than $\xi^{\perp}$. The excess current $I_{exc}$ of such point contacts usually is proportional to the order parameter near the point contact. The point contacts used by us are, however, relatively large. Their diameter is estimated to be $d^{PC}\sim 10~nm$. Allowance for the barrier reflection of electrons due to tunneling increases the values of $d^{PC}$. Consequently, the present theories used to determine this value and the temperature dependences $\Delta(T)$ and $I_{exc}(T)$ are, strictly speaking, inapplicable. We link the position of the minimum of $dV/dI$ on the $eV$ axis with the energy gap $\Delta$, noting that this minimum does not depend on the contact resistance, i.e., on the density of the current flowing through the point contact. The I-V characteristic of a point contact clearly turned out to depend on the depth at which a microscopic short circuit occurs in the superlattice. For three-layer structures ($N=1.5$), whose characteristics are shown in Figs.\ref{Fig1} and \ref{Fig2}, this short
circuit probably occurs inside the $PbTe$ layer near the superconducting heterojunction (case B in Fig.\ref{Fig1}a). This conclusion is supported by the presence of $I_{exc}$ down to the lowest temperatures because of the enrichment of the $PbTe$ layers with electrons due to the difference in the work functions of $PbTe$ and $PbS$. In some point contacts in a superlattice with $N=10$ the excess current $I_{exc}$ vanishes at $T<2.5~K$. This behavior can be explained in terms of the short-circuiting effect which occurs in an electron-depleted $PbS$ layer at a distance greater than $\xi^{\perp}(0)$ from the heterojunction. Figure \ref{Fig1}b shows the characteristics measured in various magnetic fields H oriented parallel to the layers of the superlattice, and Fig.\ref{Fig1}d is a plot of $\Delta(H)$ which was constructed for these curves under the assumption that the distance between the minima (kinks) is the quantity $2\Delta(H)$. In fields $H>15~kOe$ the gap vanishes (a gapless superconductivity).

\begin{figure}[]
\includegraphics[width=8cm,angle=0]{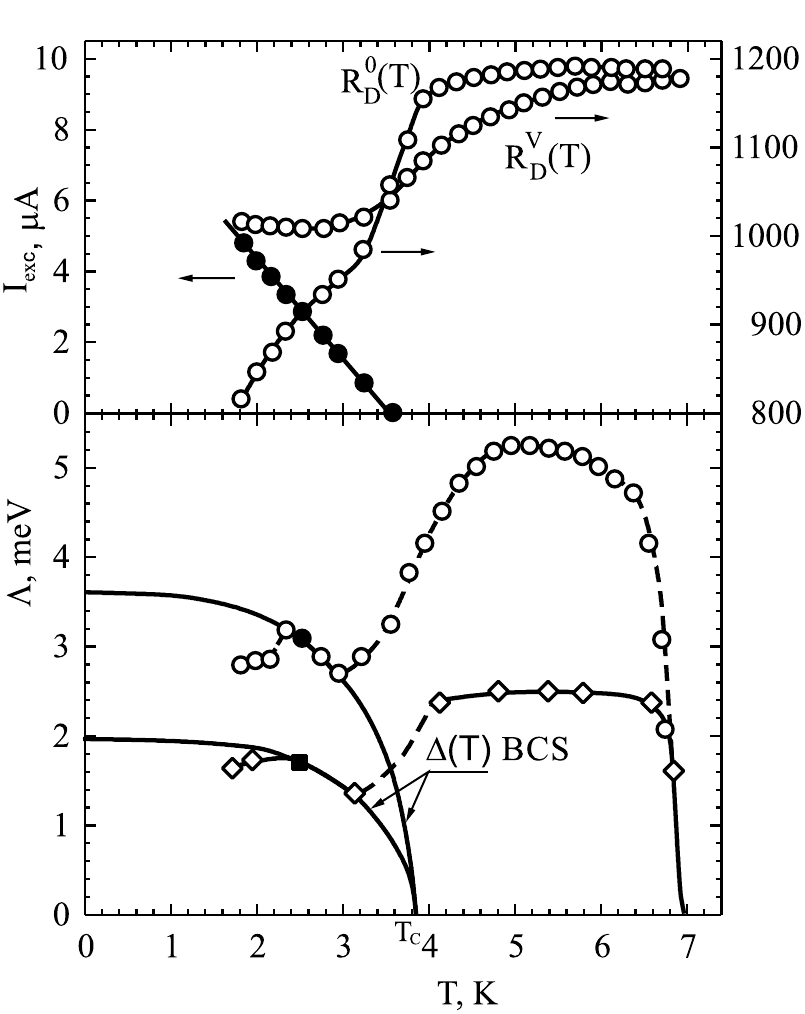}
\caption[]{Temperature dependence of the energy gap $\Delta$, of the excess current $I_{exc}$, and of the differential resistances at zero bias voltage $R_{d}(0)$ and large bias voltages $R_{d}(V)$ on a point contact whose characteristics are shown in Fig.2.\\ $\diamond$-temperature dependence of the energy gap for a superlattice with $N=10$.}
\label{Fig3}
\end{figure}

Figure \ref{Fig2} shows the families of $dV/dI$ vs $V$ and the I-V characteristics of another contact for a superlattice with $N=1.5$ at various temperatures $T$. The temperature dependences of $\Delta(T)$ for a given point contact, along with the dependence for the point contact in the superlattice with $N=10$, are shown in Fig.\ref{Fig3}. Each dependence coincides with $\Delta(T)/\Delta(0)$ of the BCS theory only in a narrow temperature interval, $2.5~Ê<T<3.2~K$. At lower temperatures the average gap in the contact region decreases since the superconductivity localizes at the grid of the misfit dislocations. At higher values of $T$, the gap corresponds to the superconductivity which localizes at a heterojunction in the same direction as the lines or the nodes of a grid of misfit dislocations. The apparent increase and spreading of the gap stem from an additional decrease of the voltage in the normal regions of the paraconducting phase. In the case of point contacts which seem to have been formed near the lower heterojunction (case C in Fig.\ref{Fig1}a), the energy gap $\Delta$ does not depend on $H$ to $30~kOe$, although $I_{exc}$
vanishes at $H>20~kOe$. For such point contacts the gap does not depend on $T$ and decreases to zero only in the immediate vicinity of $T_{c}^{*}=6-7~K$, although the superconductivity in the interior of the contact usually vanishes at $T_{c}\sim 3.5~K$, and at $T>T_{c}^{*}$ it is not observed at all. Figure \ref{Fig3} also shows the temperature dependence of the dynamic resistance for zero bias voltage $R_{d}(0)$ and for bias voltages considerably greater than $\Delta-R_{d}(V)$, where the latter quantity is nearly constant. The temperature dependence of $R_{d}(V)$ can be attributed to the change in the geometry of the current flow near the point contact. In the normal state the current flow, which is of a quasi-two-dimensional nature, is directed principally along the highly conductive heterojunction. In the region of superconducting fluctuations $R_{d}(V)$ decreases because
 of the paraconductivity which accounts for the more pronounced three-dimensional nature of the current flow. At the transition to the superconducting states near $T_{c}$, the current flow acquires even a more pronounced three-dimensional nature, and $R_{d}(V)$ becomes stabilized. Below $2.5~K$, however, $R_{d}(V)$ tends to increase as a result of localization of the superconductivity near the grid of the misfit dislocations. Because of this circumstance, the standard definition of $I_{exc}$ as the difference between the I-V characteristics of $S$ and $N$ cannot be used. We have therefore defined $I_{exc}$ as the difference between the I-V characteristics of $S$ and the line which is drawn parallel to it near the bias voltages of $15-30~meV$ and which passes through the origin of coordinates. Since $d^{PC}>\xi^{\perp}$, the temperature dependence of $I_{exc}$ corresponds to the temperature dependence of the critical current in a bulk superconductor (see Fig.86 in Ref. \cite{4}). The fact that $I_{exc}$ in the point contacts vanishes at a temperature corresponding to the loss of superconductivity in the entire sample with macroscopic dimensions suggests that the sample is perturbed only slightly during the fabrication of the point contacts. The absolute values of the gaps $\Delta(0)$ for the point contacts, whose characteristics are shown in Figs.\ref{Fig1}-\ref{Fig3}, lie in the energy interval 1.8-2.8~$meV$, which gives anomalously large values, 11-17, for the relation $\eta=2\Delta/T_c$. There are two circumstances, the account of which would reduce $\eta$. First, it is possible that the position of the minimum of $dV/dI$ corresponds to $2\Delta$, rather than to $\Delta$. Secondly, if the superconductivity along the lines of the grid of misfit dislocations is of a quasi-one-dimensional nature, the fluctuations may substantially decrease the critical temperature, and $T_{c}$ in $\eta$ should be replaced by $T_{c}^{*}\sim 7~Ê$. Both these factors have been discussed in the literature in connection with the anomalously large gaps in the quasi-one-dimensional organic superconductors \cite{5,6}. Clearly, these factors also apply to the large gaps which are usually seen in high-temperature superconductors when point-contact methods are used.

 The results of this study show that the gap may become quite large in a small, inhomogeneous superconductor adjacent to a point contact and that its temperature dependence and field dependence depend qualitatively on the position of the point contact. A similar situation apparently occurs in the case of high-$T_{c}$ superconductors. The values of $\Delta$ determined from optical measurements usually are much smaller than those from point-contact measurements since they correspond to a gap which is averaged over an area much larger than the size of the inhomogeneity.

\end{document}